\documentclass[a4paper]{jpconf}
\usepackage{graphicx}
\usepackage{xspace} 

\usepackage{ifthen}

\newboolean{pdflatex}
\setboolean{pdflatex}{true} 

\newboolean{articletitles}
\setboolean{articletitles}{true} 

\newboolean{uprightparticles}
\setboolean{uprightparticles}{false} 

\newboolean{inbibliography}
\setboolean{inbibliography}{false} 

\usepackage[top=1in, bottom=1.25in, left=1in, right=1in]{geometry}

\usepackage{microtype}
\usepackage{lineno}  
\usepackage{xspace} 
\usepackage{caption} 

\usepackage{graphicx}  
\usepackage{color}
\usepackage{colortbl}
\graphicspath{{./figs/}} 

\usepackage{amsmath} 
\usepackage{amssymb}
\usepackage{amsfonts}
\usepackage{upgreek} 

\newcommand*\patchAmsMathEnvironmentForLineno[1]{%
\expandafter\let\csname old#1\expandafter\endcsname\csname #1\endcsname
\expandafter\let\csname oldend#1\expandafter\endcsname\csname
end#1\endcsname
 \renewenvironment{#1}%
   {\linenomath\csname old#1\endcsname}%
   {\csname oldend#1\endcsname\endlinenomath}%
}
\newcommand*\patchBothAmsMathEnvironmentsForLineno[1]{%
  \patchAmsMathEnvironmentForLineno{#1}%
  \patchAmsMathEnvironmentForLineno{#1*}%
}
\AtBeginDocument{%
\patchBothAmsMathEnvironmentsForLineno{equation}%
\patchBothAmsMathEnvironmentsForLineno{align}%
\patchBothAmsMathEnvironmentsForLineno{flalign}%
\patchBothAmsMathEnvironmentsForLineno{alignat}%
\patchBothAmsMathEnvironmentsForLineno{gather}%
\patchBothAmsMathEnvironmentsForLineno{multline}%
\patchBothAmsMathEnvironmentsForLineno{eqnarray}%
}

\usepackage{hyperref}    
\usepackage[all]{hypcap} 

\usepackage{xspace} 
\usepackage{upgreek}

\def\lhcb {\mbox{LHCb}\xspace}

\def\MagUp {\mbox{\em Mag\kern -0.05em Up}\xspace}

\ifthenelse{\boolean{uprightparticles}}%
{

 \def\PDelta      {\ensuremath{\Delta}\xspace}                 
 \def\PXi      {\ensuremath{\Xi}\xspace}                 
 \def\PLambda      {\ensuremath{\Lambda}\xspace}                 
 \def\PSigma      {\ensuremath{\Sigma}\xspace}                 
 \def\POmega      {\ensuremath{\Omega}\xspace}                 
 \def\PUpsilon      {\ensuremath{\Upsilon}\xspace}                 
 
 \def\PB      {\ensuremath{\mathrm{B}}\xspace}                 
                  
 \def\PD      {\ensuremath{\mathrm{D}}\xspace}

 \def\PK      {\ensuremath{\mathrm{K}}\xspace}

 \def\Pb      {\ensuremath{\mathrm{b}}\xspace}                 
 \def\Pc      {\ensuremath{\mathrm{c}}\xspace}

 \def\Pi      {\ensuremath{\mathrm{i}}\xspace}

 \def\Pp      {\ensuremath{\mathrm{p}}\xspace}

}
{

 \mathchardef\PDelta="7101
 \mathchardef\PXi="7104
 \mathchardef\PLambda="7103
 \mathchardef\PSigma="7106
 \mathchardef\POmega="710A
 \mathchardef\PUpsilon="7107
                  
 \def\PB      {\ensuremath{B}\xspace}                 
                  
 \def\PD      {\ensuremath{D}\xspace}

 \def\PK      {\ensuremath{K}\xspace}

 \def\Pb      {\ensuremath{b}\xspace}                 
 \def\Pc      {\ensuremath{c}\xspace}

 \def\Pi      {\ensuremath{i}\xspace}

 \def\Pp      {\ensuremath{p}\xspace}

}

\makeatletter
\ifcase \@ptsize \relax
  \newcommand{\miniscule}{\@setfontsize\miniscule{4}{5}}
\or
  \newcommand{\miniscule}{\@setfontsize\miniscule{5}{6}}
\or
  \newcommand{\miniscule}{\@setfontsize\miniscule{5}{6}}
\fi
\makeatother

\DeclareRobustCommand{\optbar}[1]{\shortstack{{\miniscule (\rule[.5ex]{1.25em}{.18mm})}
  \\ [-.7ex] $#1$}}

\def\cquark    {{\ensuremath{\Pc}}\xspace}

\def\bquark    {{\ensuremath{\Pb}}\xspace}

  \def\Kbar    {{\kern 0.2em\overline{\kern -0.2em \PK}{}}\xspace}

\def\KorKbar    {\kern 0.18em\optbar{\kern -0.18em K}{}\xspace}

  \def\Dbar    {{\kern 0.2em\overline{\kern -0.2em \PD}{}}\xspace}

\def\DorDbar    {\kern 0.18em\optbar{\kern -0.18em D}{}\xspace}

\def\B       {{\ensuremath{\PB}}\xspace}
\def\Bbar    {{\ensuremath{\kern 0.18em\overline{\kern -0.18em \PB}{}}}\xspace}

\def\BorBbar    {\kern 0.18em\optbar{\kern -0.18em B}{}\xspace}

  \def\Y#1S{\ensuremath{\PUpsilon{(#1S)}}\xspace}

\def\proton      {{\ensuremath{\Pp}}\xspace}
\def\antiproton  {{\ensuremath{\overline \proton}}\xspace}

\def\Lbar        {{\ensuremath{\kern 0.1em\overline{\kern -0.1em\PLambda}}}\xspace}
\def\LorLbar    {\kern 0.18em\optbar{\kern -0.18em \PLambda}{}\xspace}

\def\to                 {\ensuremath{\rightarrow}\xspace}

\def\AT#1     {\ensuremath{A_{\mathrm{T}}^{#1}}\xspace}           

\def\C#1      {\ensuremath{\mathcal{C}_{#1}}\xspace}                       
\def\Cp#1     {\ensuremath{\mathcal{C}_{#1}^{'}}\xspace}                    
\def\Ceff#1   {\ensuremath{\mathcal{C}_{#1}^{\mathrm{(eff)}}}\xspace}        
\def\Cpeff#1  {\ensuremath{\mathcal{C}_{#1}^{'\mathrm{(eff)}}}\xspace}       
\def\Ope#1    {\ensuremath{\mathcal{O}_{#1}}\xspace}                       
\def\Opep#1   {\ensuremath{\mathcal{O}_{#1}^{'}}\xspace}                    

\newcommand{\tev}{\ifthenelse{\boolean{inbibliography}}{\ensuremath{~T\kern -0.05em eV}\xspace}{\ensuremath{\mathrm{\,Te\kern -0.1em V}}}\xspace}
\newcommand{\gev}{\ensuremath{\mathrm{\,Ge\kern -0.1em V}}\xspace}
\newcommand{\mev}{\ensuremath{\mathrm{\,Me\kern -0.1em V}}\xspace}
\newcommand{\kev}{\ensuremath{\mathrm{\,ke\kern -0.1em V}}\xspace}
\newcommand{\ev}{\ensuremath{\mathrm{\,e\kern -0.1em V}}\xspace}
\newcommand{\gevc}{\ensuremath{{\mathrm{\,Ge\kern -0.1em V\!/}c}}\xspace}
\newcommand{\mevc}{\ensuremath{{\mathrm{\,Me\kern -0.1em V\!/}c}}\xspace}
\newcommand{\gevcc}{\ensuremath{{\mathrm{\,Ge\kern -0.1em V\!/}c^2}}\xspace}
\newcommand{\gevgevcccc}{\ensuremath{{\mathrm{\,Ge\kern -0.1em V^2\!/}c^4}}\xspace}
\newcommand{\mevcc}{\ensuremath{{\mathrm{\,Me\kern -0.1em V\!/}c^2}}\xspace}

\def\invfb   {\ensuremath{\mbox{\,fb}^{-1}}\xspace}

\def\sec  {\ensuremath{\mathrm{{\,s}}}\xspace}
\def\ms   {\ensuremath{{\mathrm{ \,ms}}}\xspace}

\def\gsim{{~\raise.15em\hbox{$>$}\kern-.85em
          \lower.35em\hbox{$\sim$}~}\xspace}
\def\lsim{{~\raise.15em\hbox{$<$}\kern-.85em
          \lower.35em\hbox{$\sim$}~}\xspace}

\def\pt         {\mbox{$p_{\mathrm{ T}}$}\xspace}

\def\tell1  {TELL1\xspace}
\def\ukl1   {UKL1\xspace}

\def\Sigmaplus{\ensuremath{\PSigma^+}\xspace}

\def\sigmapmumu{\ensuremath{\PSigma^+ \to \Pp \mu^+ \mu^-}\xspace}
\def\pmumu{\ensuremath{\Pp \mu^+ \mu^-}\xspace}
\def\mpmumu{\ensuremath{m_{\Pp \mu^+ \mu^-}}\xspace}

\def\sigmapxmumu{\ensuremath{\PSigma^+ \to \Pp X^0 (\to \mu^+ \mu^-)}\xspace}
\def\sigmapmumulfv{\ensuremath{\PSigma^+ \to \antiproton \mu^+ \mu^+}\xspace}
\def\sigmappiz{\ensuremath{\PSigma^+ \to \Pp \pi^0}\xspace}
\def\sigmappizero{\sigmappiz}
\def\mppizero{\ensuremath{m_{\Pp \pi^0}}\xspace}
\def\pizero{\ensuremath{\pi^0}\xspace}

\def\kpipipi{\ensuremath{K^+ \to \pi^+ \pi^- \pi^+}\xspace}
\def\kpimumu{\ensuremath{K^+ \to \pi^+ \mu^- \mu^+}\xspace}

\def\mmumu{\ensuremath{m_{\mu^+\mu^-}}\xspace}
\def\lambdappi{\ensuremath{\PLambda^0 \to \Pp \pi^-}\xspace}
\def\B{\ensuremath{\mathcal{B}}\xspace}

\def\bujpsikstar{\ensuremath{B^+ \to J/\psi K^{\ast +}(\to K^+ \pi^0) }\xspace}
\def\bujpsik{\ensuremath{B^+ \to J/\psi K^+}\xspace}

\usepackage{booktabs}

\usepackage{xcolor} 
\definecolor{halfgray}{gray}{0.55} 
\definecolor{webgreen}{rgb}{0,.5,0}
\definecolor{webbrown}{rgb}{.6,0,0}
\definecolor{Maroon}{cmyk}{0, 0.87, 0.68, 0.32}
\definecolor{RoyalBlue}{cmyk}{1, 0.50, 0, 0}

\hypersetup{%
    colorlinks=true, linktocpage=true, pdfstartpage=3, pdfstartview=FitV,%
    breaklinks=true, pdfpagemode=UseNone, pageanchor=true, pdfpagemode=UseOutlines,%
    plainpages=false, bookmarksnumbered, bookmarksopen=true, bookmarksopenlevel=1,%
    urlcolor=webbrown, linkcolor=RoyalBlue, citecolor=webgreen, 
    pdftitle={Evidence for the rare decay Sigma+ -> p mu+ mu-  },%
    pdfauthor={Francesco Dettori},%
    pdfsubject={},%
    pdfkeywords={},%
    pdfcreator={pdfLaTeX},%
  }

\usepackage{cite} 
\usepackage{mciteplus}
\usepackage{overpic}
\usepackage{tikz}
\usetikzlibrary{decorations.pathreplacing,decorations.pathmorphing,
decorations.markings,trees,calc,patterns}
\usetikzlibrary{positioning,arrows}
\usetikzlibrary{decorations.pathmorphing}
\usetikzlibrary{decorations.markings}

\tikzset{
photon/.style={decorate, decoration={snake}, draw=red},
higgs/.style={decorate, dashed, draw=red},
particle/.style={draw=blue, postaction={decorate},decoration={markings,mark=at 
position .5 with {\arrow[draw=blue]{>}}}},
antiparticle/.style={draw=blue, 
postaction={decorate},decoration={markings,mark=at position .5 with 
{\arrow[draw=blue]{<}}}}, 
gluon/.style={decorate, draw=black,decoration={snake,amplitude=4pt, segment 
length=5pt}}, 
majorana/.style={draw=black, postaction={decorate},decoration={markings,mark=at 
position .48 with {\arrow[draw=black]{>}},mark=at position .52 with 
{\arrow[draw=black]{<}}}},
gluonloop/.style={circle, decorate, draw=black, 
decoration={coil,aspect=1.2,amplitude=2pt, segment length=4pt},minimum 
height=1.2em},
}

\def\NSigmappizero{\ensuremath{(1\,711 \pm 9)\times 10^{3}}\xspace}
\def\NTOTSigma{\ensuremath{4 \times 10^{11}}\xspace}
\def\Nkpipipi{\ensuremath{(966 \pm 2)\times 10^{3}}\xspace} 

\def\alphatis{\ensuremath{(1.1\pm 0.6)\times 10^{-8}}\xspace}

\def\expevents{\ensuremath{4.6 \pm 4.2}\xspace}

\def\signdefault{\ensuremath{4.0}\xspace}

\def\nsigmadefault{\ensuremath{12.9^{+5.1}_{-4.2}}\xspace}

\def\upperlimit{\ensuremath{6.3 \times 10^{-8}}\xspace}

\begin{document}
\title{Evidence for the rare decay \sigmapmumu at LHCb}

\author{Francesco Dettori}

\address{CERN European Organization for Nuclear Research, CH1211 Geneve 23, Switzerland}

\ead{francesco.dettori@cern.ch}

\begin{abstract}
 A search for the rare decay \sigmapmumu is performed using $pp$ collision data recorded by the LHCb experiment
at centre-of-mass energies $\sqrt{s} = 7$ and $8 \tev$, corresponding to an integrated luminosity of $3 \invfb$. 
An excess of events is observed with respect to the background expectations with a signal significance of \signdefault standard deviations.
 No significant structure is observed in the dimuon invariant mass distribution. 
 
\end{abstract}

\section{Introduction}
\label{sec:Introduction}

The \sigmapmumu decay\footnote{The inclusion of charge conjugated processes is implied throughout this note.}
is a flavour changing neutral current process, allowed only at loop level in the standard model (SM).
The process is dominated by long-distance contributions for a predicted branching fraction of 
$\mathcal{B}(\sigmapmumu) \in [1.6, 9.0] \times 10^{-8}$~\cite{He:2005yn}, 
while the short-distance SM contributions are suppressed at a branching fraction of about $10^{-12}$.
Evidence for this decay was seen by the HyperCP experiment~\cite{Park:2005eka} with a measured branching fraction 
$\mathcal{B}(\sigmapmumu) = (8.6^{+6.6}_{-5.4} \pm 5.5) \times 10^{-8}$,  which is compatible with the SM.
Remarkably, the three observed decays have almost the same dimuon pair invariant mass of
$m_{X^0}  = 214.3 \pm 0.5 \mevcc$, which lies close to the kinematic limit.
Such a distribution, if confirmed, would point towards a decay 
with an intermediate particle coming from the \Sigmaplus baryon and decaying into 
two muons, \emph{i.e.} a \sigmapxmumu decay, which would constitute evidence of physics beyond the SM (BSM). 
Various BSM explanations have been proposed to explain the HyperCP result. The intermediate $X^0$ particle 
could be for example a light pseudoscalar Higgs boson~\cite{He:2006fr,He:2006uu} in a NMSSM or 2HDM model, 
or a sgoldstino~\cite{Gorbunov:2005nu,Demidov:2006pt} in various supersymmetric models. Other interpretations and implications 
can be found in Ref.~\cite{He:2005we,Geng:2005ra,Deshpande:2005mb,Chen:2007uv,Xiangdong:2007vv,Mangano:2007gi,Pospelov:2008zw}; 
in general a pseudoscalar particle is favoured over a scalar one and a lifetime of order $10^{-14}\sec$ is estimated for the former case.
This lifetime would correspond to a prompt $X^0$ signal, decaying in the same vertex as the \Sigmaplus baryon, 
in any present search for this particle.
Attempts to confirm the presence of this $X^0$ particle have been made by different experiments in various initial and final states
without finding any signal
\cite{Love:2008aa,Tung:2008gd,Abazov:2009yi,Aubert:2009cp,Hyun:2010an,Abouzaid:2011mi,Ablikim:2011es,Lees:2014xha},
and in LHCb through the final states $B^{0}_{(s)}\to \mu^+\mu^-\mu^+\mu^-$~\cite{Aaij:2013lla} and $B^0\to K^{\ast 0}\mu^+\mu^-$~\cite{Aaij:2015tna}.
However, the search for the \sigmapmumu decay has not been repeated,
mainly due to the absence of experiments with large hyperon production and due to experimental difficulties.

Hyperons are produced copiously in high energy proton-proton collisions at the Large Hadron Collider (LHC).
A search for \sigmapmumu decays at the LHCb experiment could therefore confirm or disprove the HyperCP evidence, and measure its branching fraction. 
This search was also suggested in Ref.~\cite{Park:2010zze}.
In this report, the search for the \sigmapmumu decay is presented as performed using $pp$ collision data 
recorded by the LHCb experiment at centre-of-mass energies $\sqrt{s} = 7$ and $8 \tev$, 
corresponding to an integrated luminosity of $3 \invfb$.
The measurement described here is detailed in Ref.~\cite{LHCb-CONF-2016-013}.

\section{Detector and data samples}
\label{sec:Detector}

The \lhcb detector is a single-arm forward spectrometer covering the \mbox{pseudorapidity} range $2<\eta <5$,
designed for the study of particles containing \bquark or \cquark
quarks, and is detailed elsewhere~\cite{Alves:2008zz,LHCb-DP-2014-002}.
The online event selection is performed by a trigger, which consists of a hardware stage, based on information from the calorimeter and muon
systems, followed by two software stages, the first performing a preliminary event reconstruction
based on partial information while the second applying a full event reconstruction.
Each of the three trigger stages is divided into several parallel trigger lines dedicated to different kinds of signals.
Decay particles involved in this analysis are often not able to activate one or more trigger stages, 
owing mainly to their soft transverse momenta. 
Nevertheless, since \Sigmaplus baryons are copiously produced in the $pp$ collisions recorded by LHCb, 
the present search can be performed on the events already recorded.
In the offline processing, trigger decisions are associated with reconstructed candidates.
A trigger decision on a particular line or on the full trigger can thus be ascribed to the reconstructed candidate, the rest of the event or both;
events triggered as such are defined respectively as triggered on signal (TOS), independently of signal (TIS), or triggered on both.
The estimation of the trigger efficiency when no specific path is selected is difficult, 
therefore a different strategy is adopted here: 
all the candidates passing the selection are used in the search for \sigmapmumu decays, 
while only the TIS candidates are used to convert the event yield into a branching fraction. 
This ensures a partial cancellation of the TIS trigger efficiency between signal and normalisation channels. 
Furthermore, control channels with large statistics can be exploited to estimate the trigger efficiency 
directly on data by measuring the overlap of events which are TIS and TOS simultaneously~\cite{LHCb-DP-2012-004}.

\section{Analysis strategy and selection }
\label{sec:strategy}

After the trigger, a loose selection is applied based on geometric and kinematic variables. 
Afterwards candidates are selected by means of a multivariate selection based on a boosted decision trees algorithm~(BDT)~\cite{Breiman,AdaBoost}.
The final search datasets are obtained by rejecting the background with a cut on the BDT output and on particle identification variables. 
The signal yield is obtained from a fit to the \pmumu invariant mass and is converted 
into a branching fraction by normalising to the \sigmappizero control channel. 
To avoid experimenter bias, candidates in the signal regions were not examined until
the analysis procedure had been finalised.

\begin{figure}[!hptb]
\begin{center}
\begin{overpic}[width = 0.49 \textwidth]{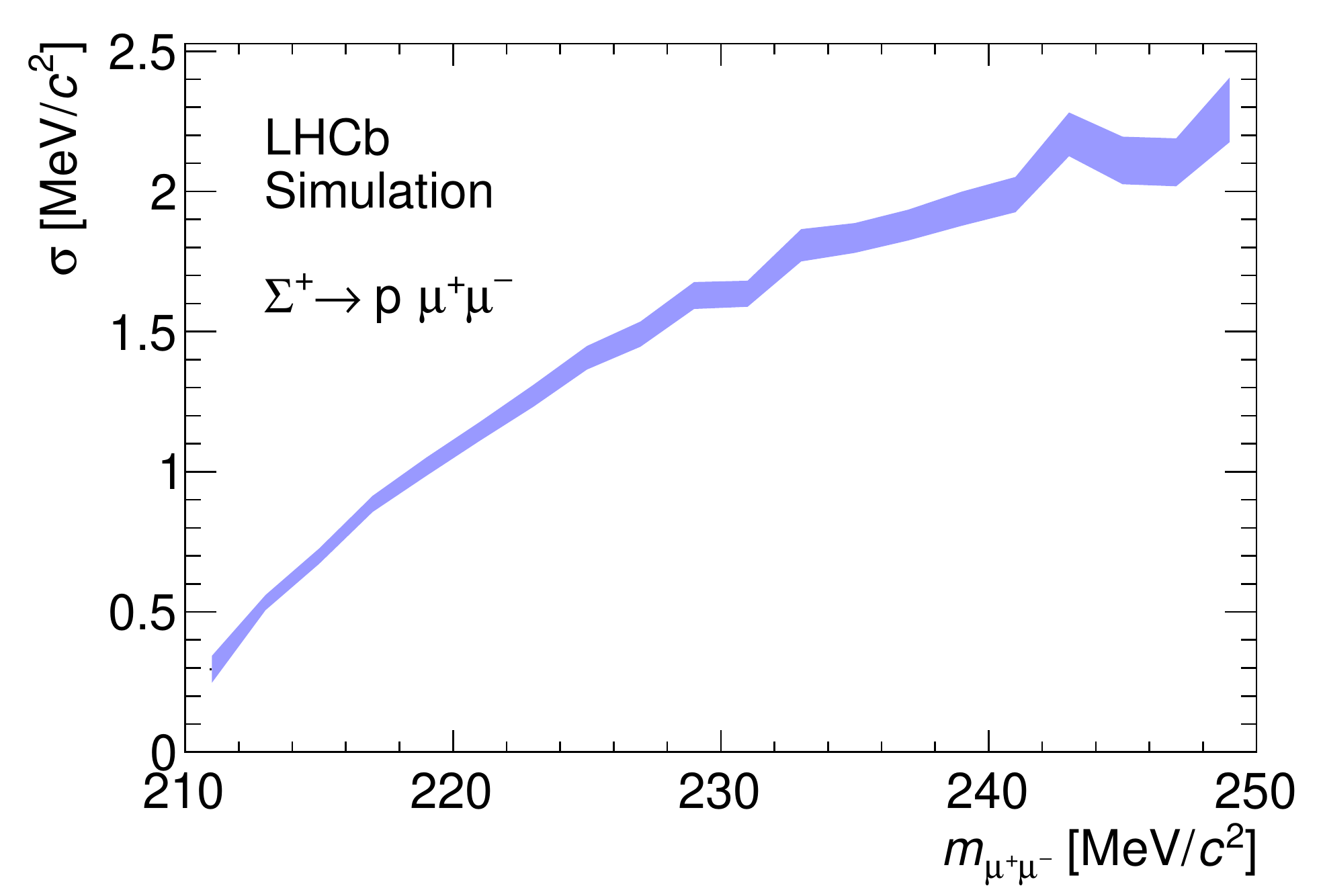}
 \put(20,15){\small (a)}
\end{overpic}
\begin{overpic}[width = 0.49 \textwidth]{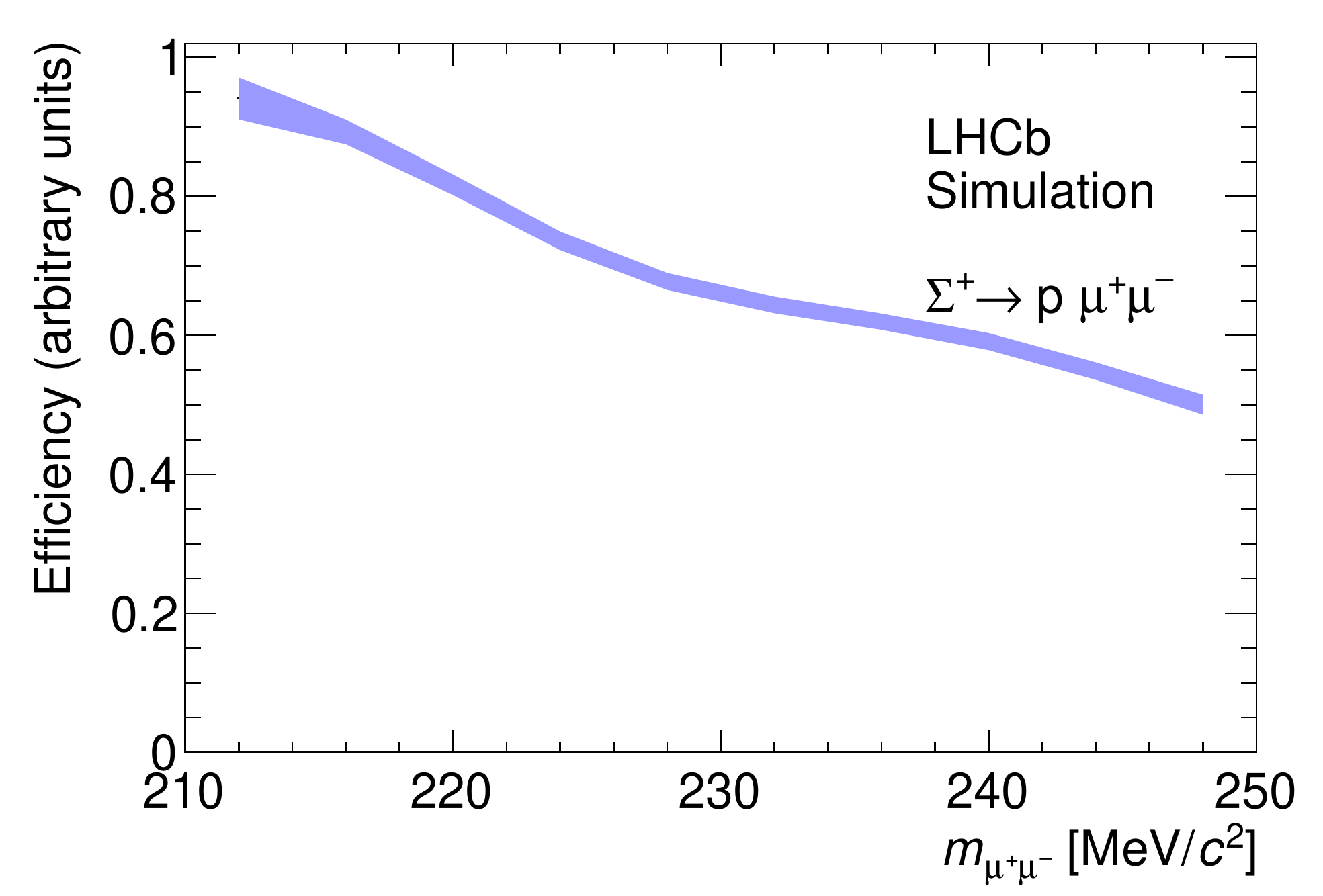}
  \put(20,15){\small (b)}
\end{overpic}
 \caption{Resolution (a)  on the dimuon invariant mass \mmumu as a function of the mass itself for simulated \sigmapmumu decays in LHCb.
 Efficiency (b) for simulated \sigmapmumu decays in LHCb as a function of the dimuon invariant mass. 
 }\label{fig:mumuperf}
 \end{center}
\end{figure}

The analysis is designed in order to search for possible peaks in the dimuon invariant mass, 
pointing towards unknown intermediate particles. 
The resolution on the dimuon invariant mass as a function of the mass itself is shown in Fig.~\ref{fig:mumuperf}(a).
The selection is devised such that no fake structures are induced in the dimuon invariant mass distribution of 
signal candidates. 
The signal efficiency varies as a function of the dimuon invariant mass and is shown in Fig.~\ref{fig:mumuperf}(b). 
The efficiency is larger at lower dimuon mass owing to the larger recoil momentum against the proton, which 
gives larger minimum transverse momentum to the two muons, ensuring a better tracking efficiency.

Together with the signal \sigmapmumu decay, the following channels are also selected in data: 
the \sigmappizero decay as normalisation channel, the \sigmapmumulfv candidate decay as control channel for the combinatorial background, 
and the \kpipipi decay as a control channel for different parts of the efficiency evaluation.
Candidate \sigmapmumu decays are selected by combining two good quality oppositely charged tracks with muon identification 
with a third track with proton identification. 
The three tracks are required to form a good quality secondary vertex (SV), displaced 
 from any $pp$ interaction vertex (PV) requiring a measured \Sigmaplus lifetime greater than 6 ps. 
Only \Sigmaplus candidates with transverse momentum $\pt > 0.5 \gevc$,  $\chi^2_{\rm IP}<36$ and 
cosine of the angle between the flight direction and the reconstructed momentum larger than 0.9 are retained. 
Candidate \sigmapmumu decays are considered only if the invariant mass satisfies $|\mpmumu - m_{\Sigmaplus}| < 500\mevcc$, 
where $m_{\Sigmaplus}$ is the known mass of the \Sigmaplus particle~\cite{PDG2014}.
A large background component is present in data due to \lambdappi decays, where the pion 
is misidentified as a muon and the two tracks are combined with a third one.
This is vetoed discarding candidates having a $p\mu^-$ pair mass within 5 \mevcc 
from the $\PLambda^0$ known mass when calculated with the $p \pi^-$ mass hypothesis.
Possible backgrounds from exclusive decays peaking in the \pmumu invariant mass have been examined, 
including \kpipipi and \kpimumu decays and various hyperon decays, and none has been found to contribute. 
In case of multiple candidates in a single event, all are retained in the selection. 
Multiple candidates are present in 5\% of the events after the initial selection, while no multiple candidate 
is present in the final selection. 
The selection for the control channel \sigmapmumulfv is identical to that of the signal but considering same-sign dimuon pairs. 
Candidate \sigmappizero decays are selected by combining one good quality track with proton identification with a \pizero reconstructed 
in the $\pizero \to \gamma \gamma$ mode from two clusters in the calorimeter. 
The selection of this decay is similar to that of the signal, but places a tighter requirement on
the proton identification, and on the transverse momenta of the daughters in order to 
reduce the high combinatorial background.
Finally, candidate \kpipipi decays  are selected from three good quality tracks, 
one of which is oppositely charged with respect to the other two; the selection is similar but tighter 
than that of the signal to cope with the high level of combinatorial background. 
See Ref.~\cite{LHCb-CONF-2016-013} for further details on the selection criteria of the different channels.

The sample of \sigmapmumu candidates in data after the selection 
is dominated by combinatorial background, part of which is due to misidentified particles.  
This is rejected by cuts on the BDT and on multivariate particle identification variables~\cite{LHCb-DP-2014-002} on the muons and on the proton. 
The BDT variable combines geometric and kinematic variables chosen so that the dependence on 
the \pmumu invariant mass and on the dimuon invariant mass is linear and small to avoid biases.
The BDT is optimised using simulated samples of \sigmapmumu events for the signal 
and candidates from the sidebands of the \sigmapmumulfv selection in data for the background.
The final cut values were chosen in order to optimise the sensitivity to a signal evidence for the 
smallest possible branching fraction~\cite{Punzi:2003bu}. 
No BDT selection is applied to the normalisation and control channels.

\section{Normalisation}
\label{sec:normalisation}

The number of signal candidates in the TIS sample is converted into a branching fraction with the formula
\begin{eqnarray*}\label{eq:normsigma}
 \B(\sigmapmumu) =
\frac{\varepsilon_{\sigmappizero}}{\varepsilon_{\sigmapmumu}}
\frac{N_{\sigmapmumu}}{N_{\sigmappizero}} \B(\sigmappizero)
 = \alpha \cdot N_{\sigmapmumu}\quad,
\end{eqnarray*}
where $\varepsilon$, $N$ and $\B$ are the efficiency, candidate yield and branching fraction of the corresponding channel, respectively,
and $\alpha$ is the single event sensitivity.

The ratio of signal and normalisation channel efficiencies, which includes the acceptance, 
the reconstruction efficiency of the final state 
particles and the selection efficiency, is computed with samples of simulated events corrected to take into account 
known differences between data and simulation.  
The reconstruction efficiency for the \pizero is calibrated using the 
ratio of \bujpsikstar and \bujpsik decays reconstructed in data.
The particle identification efficiency of protons and muons is calibrated exploiting control channels in data. 
Residual differences between data and simulation are treated as sources of systematic uncertainty. 
The signal and normalisation channels are required to be TIS at all trigger levels. 
The trigger efficiency ratio is thus expected to be unity. 
However, small differences in the average kinematic of the rest of the event 
in the two samples are present which cause the ratio to be different. 
The ratio of trigger efficiencies is thus evaluated with data-driven techniques~\cite{LHCb-DP-2012-004} 
exploiting the large sample of \kpipipi decays. 
A systematic uncertainty is assigned for the applicability of this method to the relatively soft events of this analysis.

\begin{figure}[!hptb]
 \begin{center}
\includegraphics[width = 0.49 \textwidth]{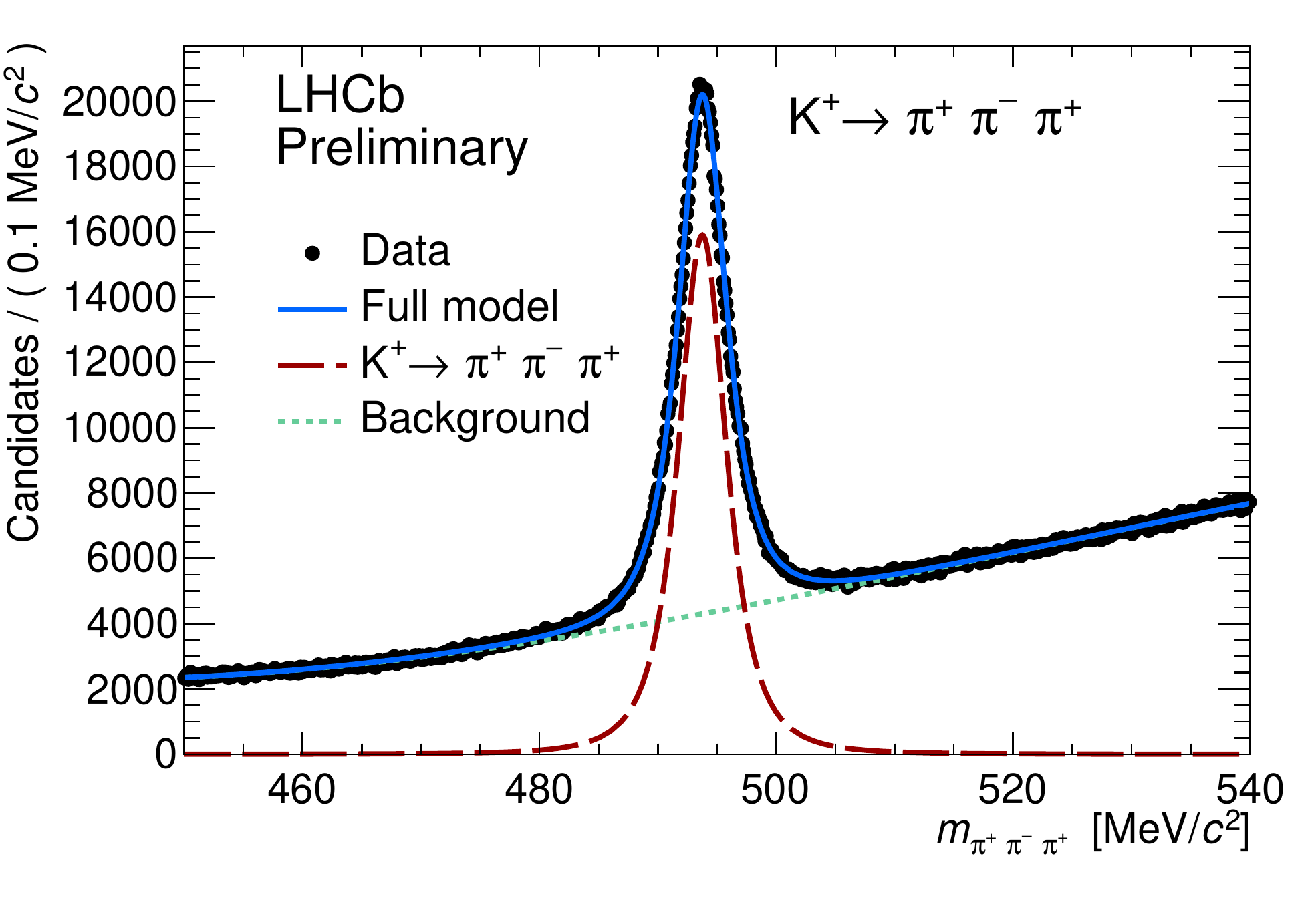}
\includegraphics[width = 0.49\textwidth]{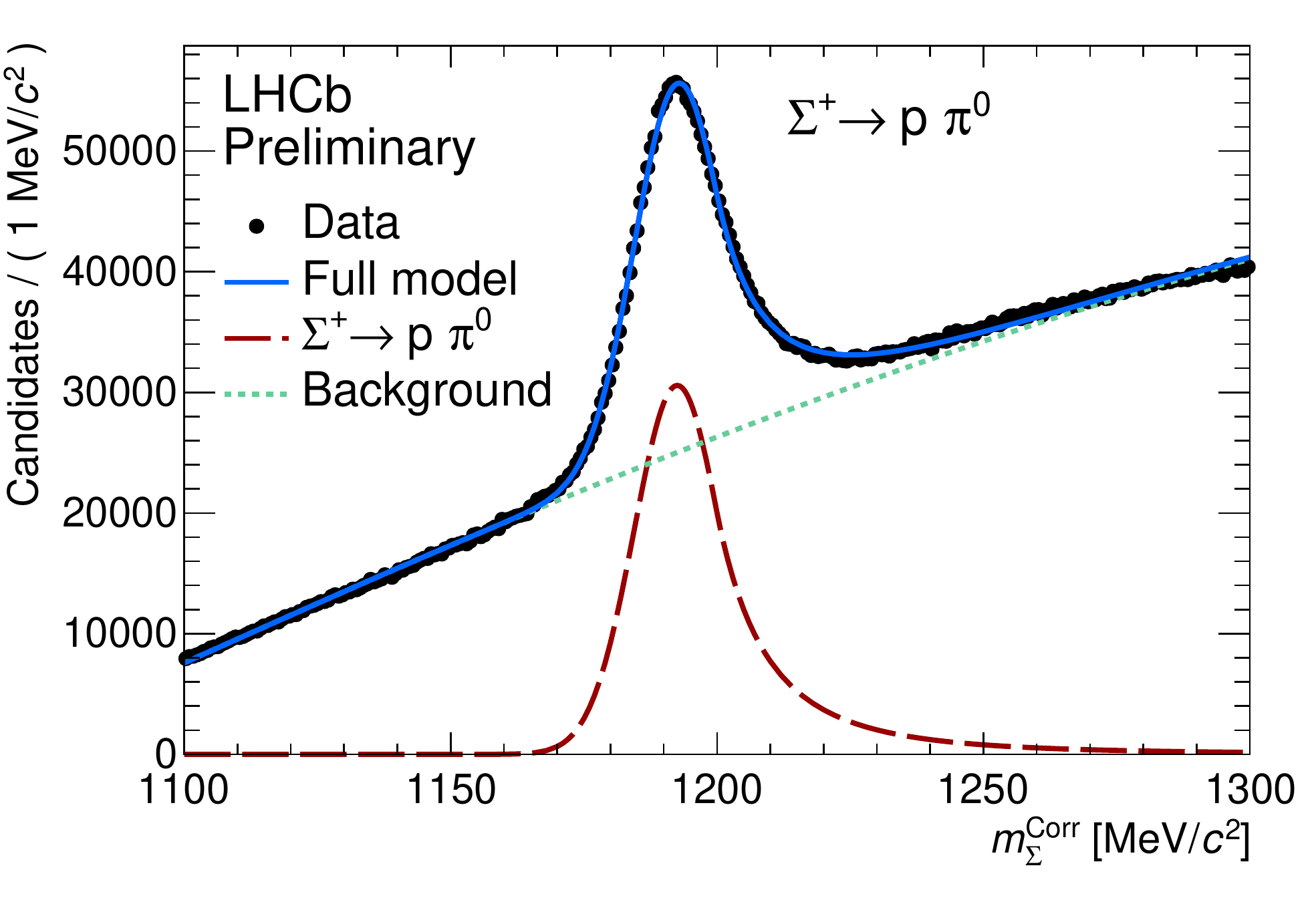}
\caption{Invariant mass distribution of (a) \kpipipi candidates superimposed with the fit to data; (b) 
fit to the distribution of the corrected mass $m_{\PSigma}^{\text{corr}}$ for \sigmappizero candidates superimposed with the fit to data. }\label{fig:masses}
 \end{center}
\end{figure}

The invariant mass distribution of the \kpipipi control channel candidates in data is shown in Fig.~\ref{fig:masses}(a).
A binned maximum likelihood extended fit is performed to the invariant mass distribution. 
The signal is described as an Hypatia function~\cite{Santos:2013gra} while the background is described 
by a second-order polynomial.  A total of \Nkpipipi \kpipipi candidates is measured.

The observed number of \sigmappizero candidates is \NSigmappizero as obtained from a binned maximum likelihood extended 
fit to the corrected invariant mass distribution.  
The corrected invariant mass is defined as $m_{\PSigma}^{\text{corr}} = \mppizero - m_{\pi^0} + m_{\pi^0}^{\text{PDG}}$ 
to correct for the \pizero mass reconstructed from the two photons. 
The \sigmappizero distribution is described as a Gaussian function with a power tail on the right side, while the background is described
by a modified ARGUS function~\cite{Albrecht:1990am}.
The invariant mass distribution is shown in Fig.~\ref{fig:masses}(b), superimposed with the fit. 
The single event sensitivity is $\alpha = \alphatis$, where the uncertainty is dominated by the aforementioned systematic uncertainties,
and corresponding to about \NTOTSigma \Sigmaplus particles produced in the LHCb acceptance in the full dataset in TIS events.
This corresponds to \expevents \sigmapmumu candidates assuming a branching fraction
of $(5\pm 4)\times 10^{-8}$, to cover the SM predicted range.

The observed number of signal \sigmapmumu events is obtained with a fit to the \pmumu invariant mass distribution 
in the range $1149.6 < \mpmumu < 1409.6 \mevcc$.
The signal distribution is described by an Hypatia function~\cite{Santos:2013gra}. 
The mass resolution and scale are calibrated using the control channel \kpipipi and comparing data and simulation distributions. 
No bias is seen in the peak position, while a 25\% correction to the resolution has to be applied to match the one observed in data. 
A resolution of $4.28 \pm 0.19 \mevcc$ is used as width of the signal \sigmapmumu distribution. 
The resolution is allowed to vary in the fit but constrained to the central value with a Gaussian constraint. 
The combinatorial background is described as a modified ARGUS function, with all parameters left free with exception of the 
threshold which is fixed to the kinematic limit. 

While the normalisation is available for the TIS sample, further studies are needed to 
calculate the trigger efficiency for the full sample. Since the TIS dataset is a sub-sample of the full sample, 
the single event sensitivity of the latter will be equal or better than the one of the former.

\section{Results and discussion}
\label{sec:results}

\begin{figure}[!htbp]
\begin{center}
\begin{overpic}[width = 0.5\textwidth]{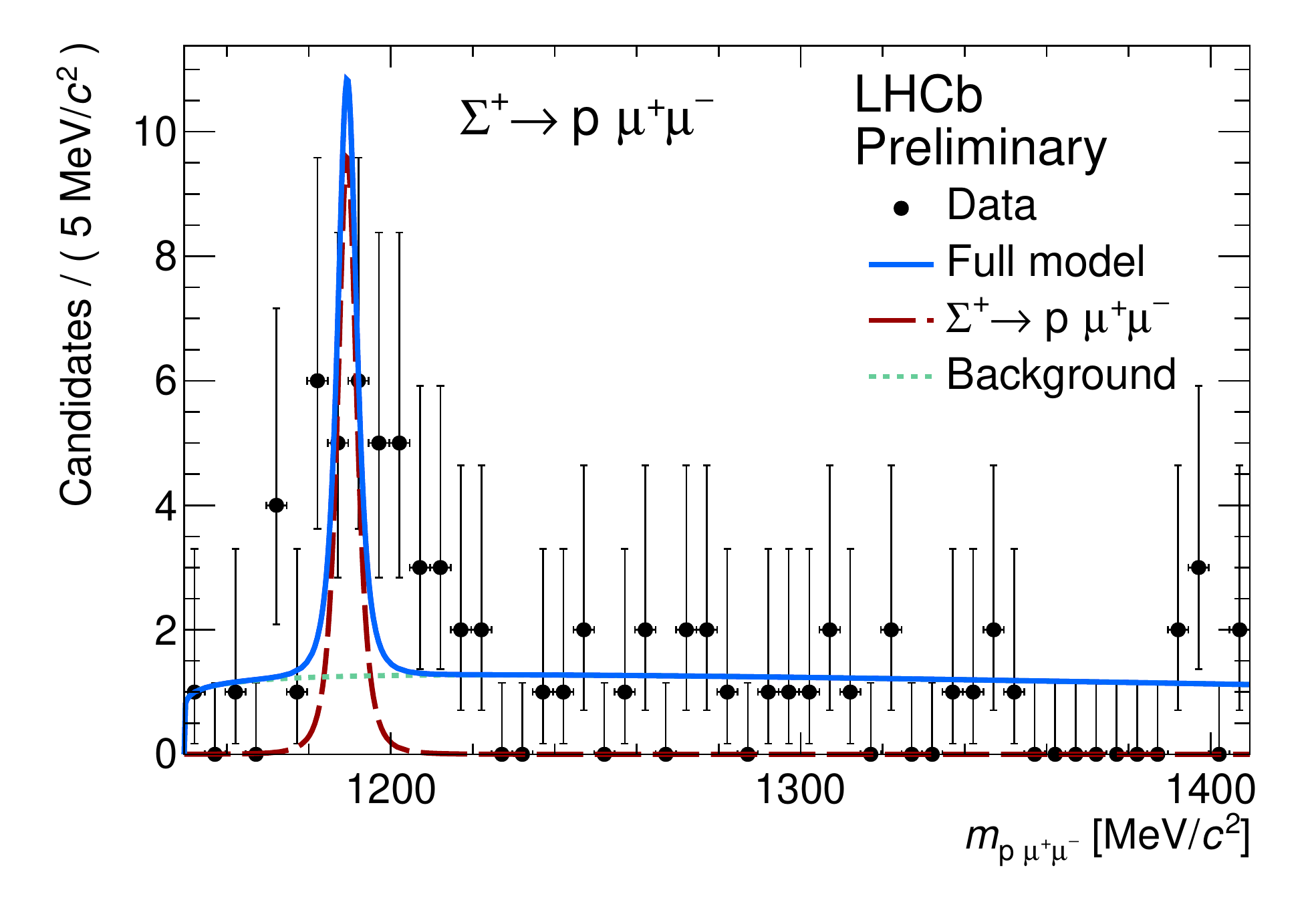}
 \put(18,57){\small (a)}
\end{overpic}\begin{overpic}[width = 0.5 \textwidth]{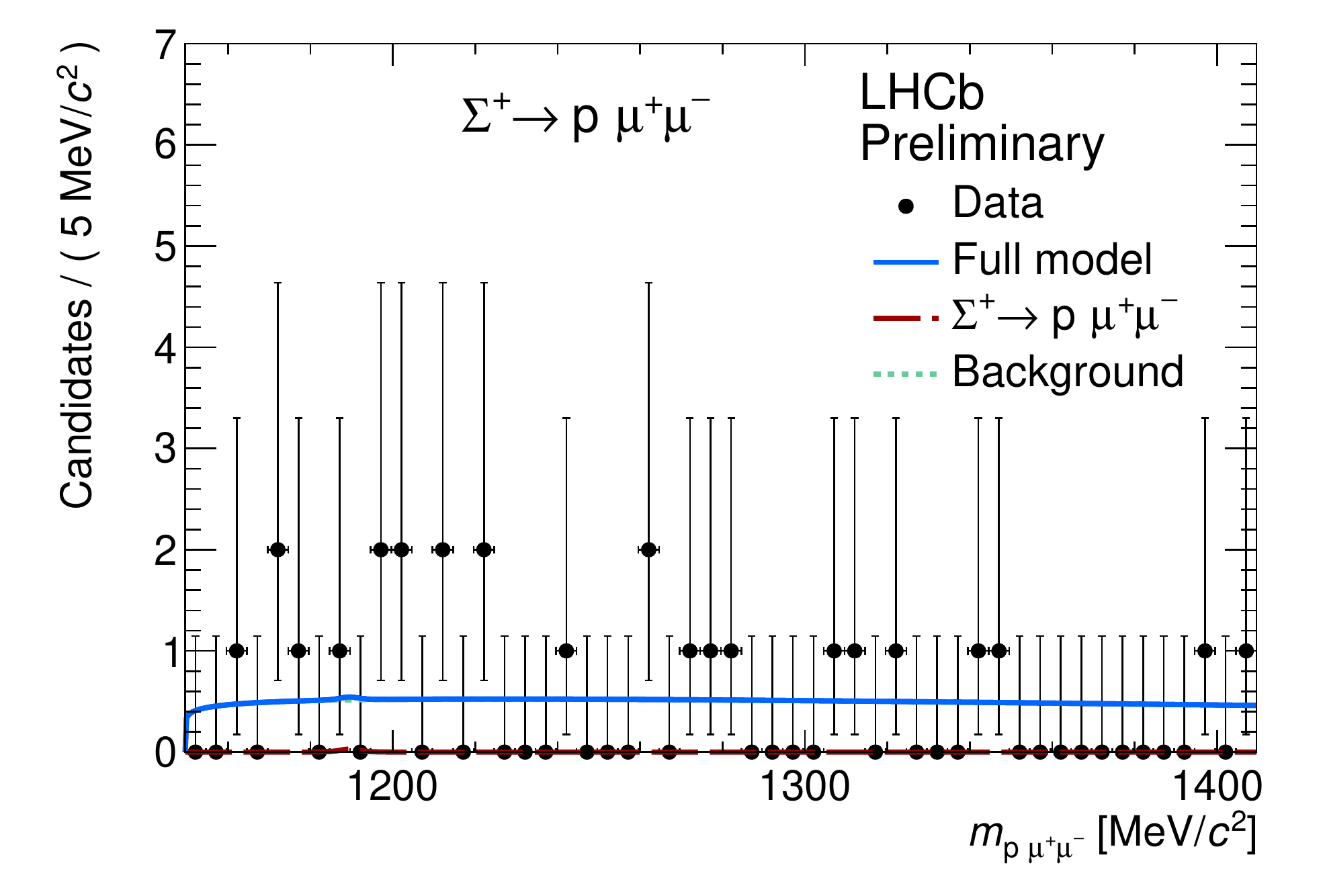}
 \put(18,57){\small (b)}
\end{overpic}
\caption{Invariant mass distribution for \sigmapmumu candidates: (a) all the candidates and (b) only the candidates 
satisfying the TIS requirement. }\label{fig:invmass}
 \end{center}
\end{figure}

The invariant mass distribution of the \sigmapmumu candidates in data is shown in Fig.~\ref{fig:invmass}(a) and (b)
for the full and TIS datasets, respectively. 
A significant signal is present in the full dataset.
The significance of the signal is of $\signdefault\,\sigma$,
obtained from the comparison of the likelihood value of the full fit with that of the background-only fit, 
and includes the relevant systematic uncertainties as gaussian constrains to the likelihood. A total of \nsigmadefault signal candidates is observed.
The signal in the TIS sample is found not to be significant.
This is due to the observed signal candidates being needed for the related event to switch on at least one of the LHCb trigger stages.
From the absence of a significant signal in the TIS dataset
an upper limit on the \sigmapmumu branching fraction 
is derived, using the CLs method~\cite{Read:2002hq}, of $\mathcal{B}(\sigmapmumu) < \upperlimit $ at 95\% confidence level (CL).

\begin{figure}[!htbp]
  \begin{overpic}[width = 0.49\textwidth]{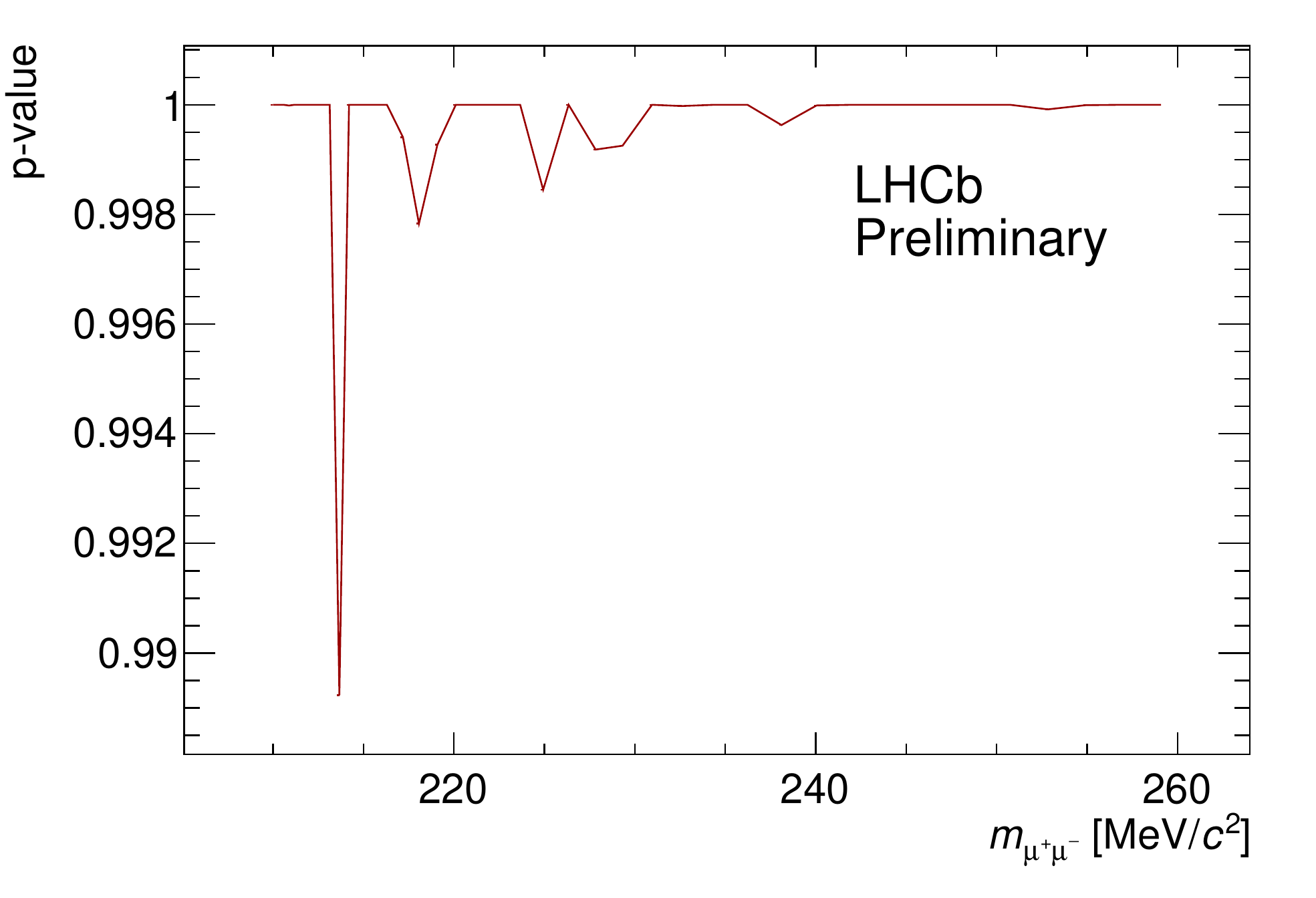}
   \put(70,45){\small (a)}
  \end{overpic}
  \begin{overpic}[width = 0.49\textwidth]{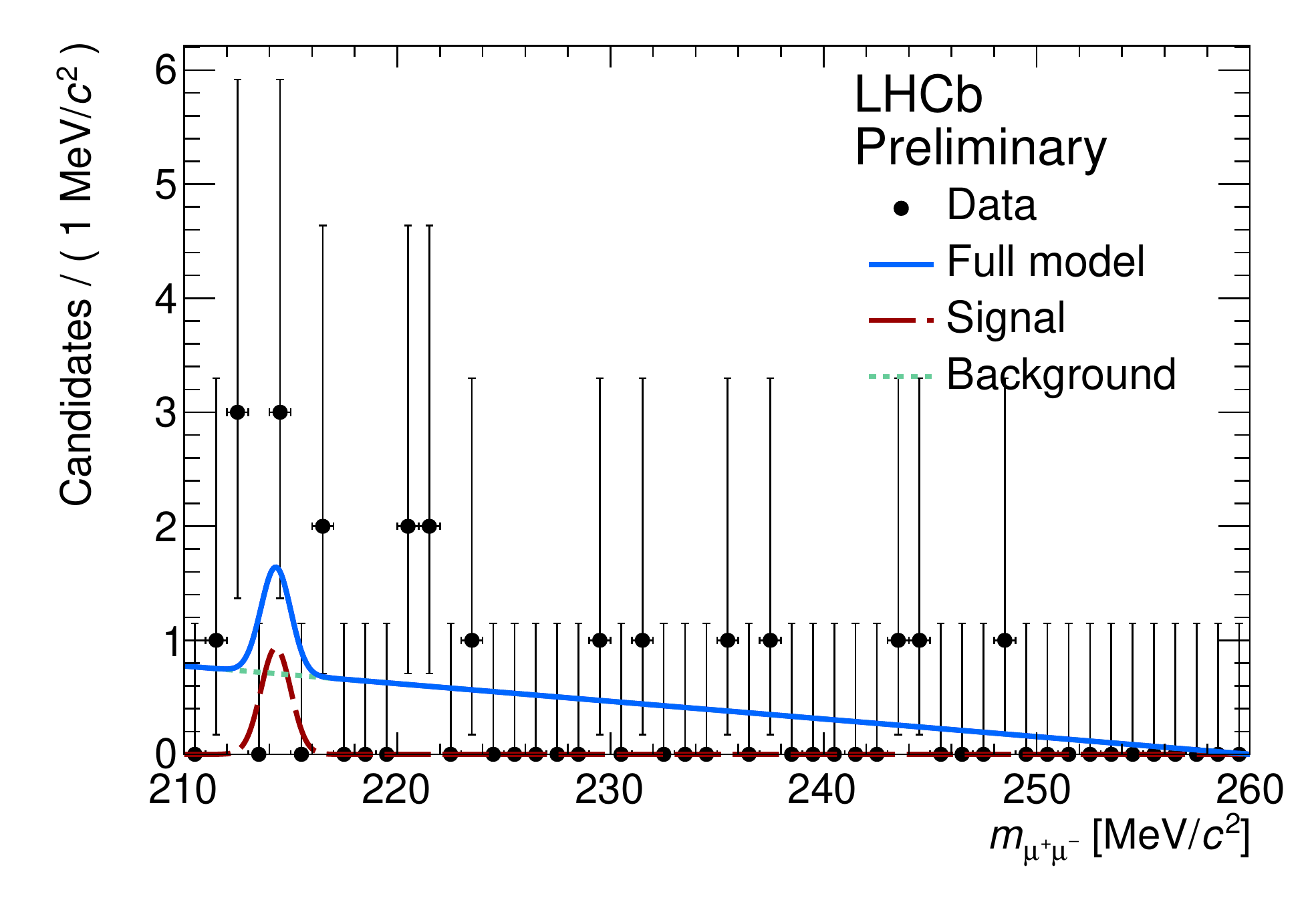}
   \put(50,45){\small (b)}
  \end{overpic}
  \caption{Plot of the local p-value as a function of the dimuon mass 
  as obtained from the fits (see text) in (a) and (b) fit for a signal with mass $m_{X^0} = 214.3\mevcc$. }
  \label{fig:mumuscan}  
 \end{figure}

Considering candidates in the full selection, a scan for possible signals 
in the dimuon invariant mass is performed, restricted to within two times the resolution
in the \sigmapmumu invariant mass around the known \Sigmaplus mass. 
The scan is performed considering a single Gaussian function for a putative $X^0$ resonance (signal) 
and a linear distribution for the remaining candidates (background). 
In Fig.~\ref{fig:mumuscan}(a) the local p-value of the background-only hypothesis as a function of the dimuon mass is shown.
No significant signal is found. 
In Fig.~\ref{fig:mumuscan}(b) the fit corresponding to a mass of $214.3\mevcc$ is shown. 
The fitted number of events for this hypothesis is $1.6 \pm 1.9$, corresponding to a fraction $0.078 \pm 0.092$ of the considered
candidates.

In summary, a search for the \sigmapmumu rare decay is performed by the LHCb experiment 
at centre-of-mass energies $\sqrt{s} = 7$ and $8 \tev$, corresponding to an integrated luminosity of $3 \invfb$.
Evidence for the \sigmapmumu decay is found in the full dataset, 
albeit not found when requiring the events to be triggered independently of the signal decay. 
The observed signal candidates show a dimuon invariant mass distribution consistent with phase space; 
no significant peak consistent with an intermediate particle is found in the dimuon 
invariant mass distribution. 
The upper limit on the branching fraction of the \sigmapmumu decay 
is \upperlimit at 95\% CL for a SM-like signal.

\section*{References}
\bibliographystyle{iopart-num}

\bibliography{main}

\providecommand{\newblock}{}
\begin{thebibliography}{10}
\expandafter\ifx\csname url\endcsname\relax
  \def\url#1{{\tt #1}}\fi
\expandafter\ifx\csname urlprefix\endcsname\relax\def\urlprefix{URL }\fi
\providecommand{\eprint}[2][]{\url{#2}}

\bibitem{He:2005yn}
He X~G, Tandean J and Valencia G 2005 {\em Phys.Rev.\/} {\bf D72} 074003
  (\textit{Preprint} \eprint{hep-ph/0506067})

\bibitem{Park:2005eka}
Park H {\em et~al.\/} (HyperCP collaboration) 2005 {\em Phys.Rev.Lett.\/} {\bf
  94} 021801 (\textit{Preprint} \eprint{hep-ex/0501014})

\bibitem{He:2006fr}
He X~G, Tandean J and Valencia G 2007 {\em Phys.Rev.Lett.\/} {\bf 98} 081802
  (\textit{Preprint} \eprint{hep-ph/0610362})

\bibitem{He:2006uu}
He X~G, Tandean J and Valencia G 2006 {\em Phys. Rev.\/} {\bf D74} 115015
  (\textit{Preprint} \eprint{hep-ph/0610274})

\bibitem{Gorbunov:2005nu}
Gorbunov D and Rubakov V 2006 {\em Phys.Rev.\/} {\bf D73} 035002
  (\textit{Preprint} \eprint{hep-ph/0509147})

\bibitem{Demidov:2006pt}
Demidov S~V and Gorbunov D~S 2007 {\em JETP Lett.\/} {\bf 84} 479--484
  (\textit{Preprint} \eprint{hep-ph/0610066})

\bibitem{He:2005we}
He X~G, Tandean J and Valencia G 2005 {\em Phys.Lett.\/} {\bf B631} 100--108
  (\textit{Preprint} \eprint{hep-ph/0509041})

\bibitem{Geng:2005ra}
Geng C~Q and Hsiao Y~K 2006 {\em Phys. Lett.\/} {\bf B632} 215--218
  (\textit{Preprint} \eprint{hep-ph/0509175})

\bibitem{Deshpande:2005mb}
Deshpande N, Eilam G and Jiang J 2006 {\em Phys.Lett.\/} {\bf B632} 212--214
  (\textit{Preprint} \eprint{hep-ph/0509081})

\bibitem{Chen:2007uv}
Chen C~H, Geng C~Q and Kao C~W 2008 {\em Phys.Lett.\/} {\bf B663} 400--404
  (\textit{Preprint} \eprint{0708.0937})

\bibitem{Xiangdong:2007vv}
Xiangdong G, Li C~S, Li Z and Zhang H 2008 {\em Eur.Phys.J.\/} {\bf C55}
  317--324 (\textit{Preprint} \eprint{0712.0257})

\bibitem{Mangano:2007gi}
Mangano M~L and Nason P 2007 {\em Mod.Phys.Lett.\/} {\bf A22} 1373--1380
  (\textit{Preprint} \eprint{0704.1719})

\bibitem{Pospelov:2008zw}
Pospelov M 2009 {\em Phys. Rev.\/} {\bf D80} 095002 (\textit{Preprint}
  \eprint{0811.1030})

\bibitem{Love:2008aa}
Love W {\em et~al.\/} (CLEO collaboration) 2008 {\em Phys. Rev. Lett.\/} {\bf
  101} 151802 (\textit{Preprint} \eprint{0807.1427})

\bibitem{Tung:2008gd}
Tung Y~C {\em et~al.\/} (E391a collaboration) 2009 {\em Phys. Rev. Lett.\/}
  {\bf 102} 051802 (\textit{Preprint} \eprint{0810.4222})

\bibitem{Abazov:2009yi}
Abazov V~M {\em et~al.\/} (D0 collaboration) 2009 {\em Phys. Rev. Lett.\/} {\bf
  103} 061801 (\textit{Preprint} \eprint{0905.3381})

\bibitem{Aubert:2009cp}
Aubert B {\em et~al.\/} (BaBar collaboration) 2009 {\em Phys. Rev. Lett.\/}
  {\bf 103} 081803 (\textit{Preprint} \eprint{0905.4539})

\bibitem{Hyun:2010an}
Hyun H~J {\em et~al.\/} (Belle collaboration) 2010 {\em Phys. Rev. Lett.\/}
  {\bf 105} 091801 (\textit{Preprint} \eprint{1005.1450})

\bibitem{Abouzaid:2011mi}
Abouzaid E {\em et~al.\/} (KTeV collaboration) 2011 {\em Phys. Rev. Lett.\/}
  {\bf 107} 201803 (\textit{Preprint} \eprint{1105.4800})

\bibitem{Ablikim:2011es}
Ablikim M {\em et~al.\/} (BESIII collaboration) 2012 {\em Phys. Rev.\/} {\bf
  D85} 092012 (\textit{Preprint} \eprint{1111.2112})

\bibitem{Lees:2014xha}
Lees J~P {\em et~al.\/} (BaBar collaboration) 2014 {\em Phys. Rev. Lett.\/}
  {\bf 113} 201801 (\textit{Preprint} \eprint{1406.2980})

\bibitem{Aaij:2013lla}
Aaij R {\em et~al.\/} (LHCb collaboration) 2013 {\em Phys. Rev. Lett.\/} {\bf
  110} 211801 (\textit{Preprint} \eprint{1303.1092})

\bibitem{Aaij:2015tna}
Aaij R {\em et~al.\/} (LHCb collaboration) 2015 {\em Phys. Rev. Lett.\/} {\bf
  115} 161802 (\textit{Preprint} \eprint{1508.04094})

\bibitem{Park:2010zze}
Park H~K 2010 {\em JHEP\/} {\bf 10} 052

\bibitem{LHCb-CONF-2016-013}
Alves~Jr A~A {\em et~al.\/} (LHCb Collaboration) 2016 {Evidence for the rare
  decay $\Sigma^+ \to p \mu^+ \mu^-$} {LHCb-CONF-2016-013}
  \urlprefix\url{https://cds.cern.ch/record/2224468}

\bibitem{Alves:2008zz}
Alves~Jr A~A {\em et~al.\/} (LHCb collaboration) 2008 {\em JINST\/} {\bf 3}
  S08005

\bibitem{LHCb-DP-2014-002}
Aaij R {\em et~al.\/} (LHCb collaboration) 2015 {\em Int. J. Mod. Phys.\/} {\bf
  A30} 1530022 (\textit{Preprint} \eprint{1412.6352})

\bibitem{LHCb-DP-2012-004}
Aaij R {\em et~al.\/} 2013 {\em JINST\/} {\bf 8} P04022 (\textit{Preprint}
  \eprint{1211.3055})

\bibitem{Breiman}
Breiman L, Friedman J~H, Olshen R~A and Stone C~J 1984 {\em Classification and
  regression trees\/} (Belmont, California, USA: Wadsworth international group)

\bibitem{AdaBoost}
Schapire R~E and Freund Y 1997 {\em J. Comput. Syst. Sci.\/} {\bf 55} 119

\bibitem{PDG2014}
Olive K~A {\em et~al.\/} (Particle Data Group) 2014 {\em Chin. Phys.\/} {\bf
  C38} 090001 {and 2015 update}

\bibitem{Punzi:2003bu}
Punzi G 2003 {\em Statistical Problems in Particle Physics, Astrophysics, and
  Cosmology\/} ed {Lyons} L, {Mount} R and {Reitmeyer} R p~79
  (\textit{Preprint} \eprint{physics/0308063})

\bibitem{Santos:2013gra}
Martinez~Santos D and Dupertuis F 2014 {\em Nucl. Instrum. Meth.\/} {\bf A764}
  150--155 (\textit{Preprint} \eprint{1312.5000})

\bibitem{Albrecht:1990am}
Albrecht H {\em et~al.\/} (ARGUS collaboration) 1990 {\em Phys. Lett.\/} {\bf
  B241} 278--282

\bibitem{Read:2002hq}
Read A~L 2002 {\em J. Phys.\/} {\bf G28} 2693--2704

\end{thebibliography}

\end{document}